# Load Balancing Strategies to Solve Flowshop Scheduling on Parallel Computing

Zheng SUN, Xiaohong HUANG, and Yan MA

School of Information Engineering and School of Computer Sciences

Beijing University of Posts and Telecommunications, Beijing, China

zhengs.bupt@gmail.com

*Abstract*—This paper first presents a parallel solution for the Flowshop Scheduling Problem in parallel environment, and then proposes a novel load balancing strategy. The proposed Proportional Fairness Strategy (PFS) takes computational performance of computing process sets into account, and assigns additional load to computing nodes proportionally to their evaluated performance. In order to efficiently utilize the power of parallel resource, we also discuss the data structure used in communications among computational nodes and design an optimized data transfer strategy. This data transfer strategy combined with the proposed load balancing strategy have been implemented and tested on a super computer consisted of 86 CPUs using MPI as the middleware. The results show that the proposed PFS achieves better performance in terms of computing time than the existing Adaptive Contracting Within Neighborhood Strategy. We also show that the combination of both the Proportional Fairness Strategy and the proposed data transferring strategy achieves additional 13~15% improvement in efficiency of parallelism.

## I. INTRODUCTION

Parallel computing is seemed as a promising solution of extremely complex problems, such as the NP hard problem. Many works, such as [2] and [4], have been presented focusing on applications of high complexity problems using parallel computing resources. Many researchers have also dedicated themselves to the improvement of parallel efficiency. In [5], the author shows that hierarchical master-worker scheme outperforms conventional ones in avoidance of communication congestions. In [3], the author proves that the well known Adaptive Contracting within Neighborhood Strategy (ACWN) can achieve a better load distribution than the Randomized Allocation Strategy (RAND) [13] in general tree-structured computations. In later part of this paper, we will show that the proposed Proportional Fairness Strategy can lead to a better performance than ACWN.

The classical Flowshop Scheduling Problem has been studied for decades. Reference [1] provides a statistical review of the literature in this specific topic. So far, many academic contributions have focused on the improvement of heuristic algorithms, like [8] [10].

Besides those focusing on sequential computation, there are also contributions which present parallel solutions for this problem using computing grid resources. In [2] the author manages to provide a hybrid approach which combines multiple algorithms to improve efficiency. Works like [2] mostly focus on the invention of more powerful algorithms for achieving efficient utility of computing resources. But technologies of load distribution and process mapping are not discussed in monographs.

This paper aims to present an implementation for parallel solution of Flowshop Scheduling Problem, and meanwhile, propose a new load distribution strategy to effectively run the NP problems on the super computer.

The Branch and bound algorithm is widely used to solve optimization problems. However, many of these attempts tend to provide efficiency among computing resources by steal/allocate the identical amount of tasks from/to nodes. In the situations where the fluctuation of network connection quality exists, the efficiency of this strategy falls down inevitably. In order to solve this problem, the Proportional Fairness Strategy is used in this paper to achieve load balancing. The idea of this strategy is quite straightforward. When the master node is about to balance system load, the computational performance of nodes in its processing sets is taken into account. Therefore, the tasks reallocated in every node will be examined so that more tasks would be assigned to the processing sets with better computational conditions. For data transfer strategy, it also shows that the data type used in communications could be improved so that the bytes transferred are reduced from nodes to nodes.

The application of this paper is parallelized under the hierarchical master-worker paradigm described in [11], and implemented on a super computer consisting of 86 CPUs, in which the middleware used is MPICH [12]. During the evaluation, we choose 10 cases from the Taillard's benchmarks and several randomly benchmarks are generated to test and compare the performance of every proposed algorithm.

The rest of this paper is organized as follow. Section 2 shows the parallelization and the design of internodes communications. Section 3 describes and compares the ACWN and the PFS load distribution strategies. Section 4 presents data transfer optimization. Section 5 shows experimental results. Finally, Section 6 gives conclusions and future works.

This research is sponsored by Project 60772108 supported by National Natural Science Foundation of China and National Basic Research Program of China (973 Program), 2007CB310604

## II. Parallel Method to Find The Best Global Solution

The Branch and Bound Algorithm is the core of a sequential exhaustive method, and it is able to be parallelized on distributed networks by deploying the computation of subproblems on computing nodes. The paradigms used in parallel branch and bound algorithm have been proposed in many literatures, like [2] [4].

The simplest method to implement parallel distribution is to replicate the whole program in every node and let them take responsibility to compute only a tiny part of the problem. This method could achieve the maximum parallel acceleration rate in principle. However, due to the particular feature of the branch and bound algorithm, the computation load in every subproblem is not the same, and may even have a considerable fluctuation. In this simple strategy, since there are rare communications between nodes, if a better local solution is found in a node, it is impossible to broadcast this valuable information to other nodes promptly, thus causes the decrease of efficiency due to the delay of prunes of redundant branches.

Rather, we take benchmark variances into consideration, and adopt the hierarchical master-worker system, which is shown in Figure 1. It can greatly avoid performance degradation [11].

The hierarchical master-worker system is derived from traditional master-worker paradigm, but differentiates master processes from supervisor processes, which control multiple computing process sets containing a single master process and multiple worker processes. Distributed tasks are delivered from supervisor processes to master processes then to worker processes hierarchically. The collection of computed results is performed in the reverse way. The task balance is performed in the supervisors in several schemes, which will be discussed in section 3.

During the research based on this paper, a two level hierarchical Master-Slave paradigm has been employed. Since the time span of computing a branch is flexible, different amount of tasks should be allocated to different nodes according to its completion. Therefore, every node would ask others if there is a better solution has been found. If so, it requests the new solution to replace the old one it stores.

In the whole scenario, the supervisor processes are rare, which take responsibility of balancing the task allocation and ensure every worker node has the most up-to-date release of the local best solution. Here are task descriptions of a supervisor except relating to task balance.

➢ Requests the current local best solutions from every master node, and compares them.

➢ Records the best solution received from a master node and sends it to other master nodes to refresh.

The master nodes not being responsible for computing is with tasks listed as follows:

➢ Check the current local best solution in the supervisor node as soon as getting reception of the corresponding requests.

➢ Record the current local best solution found by the worker nodes.

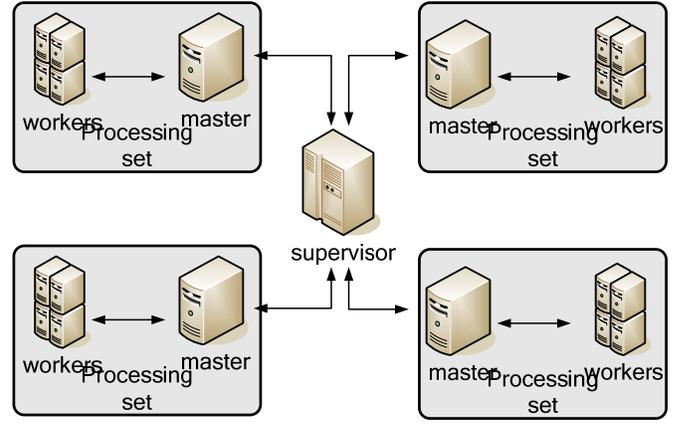

Fig. 1. The topography of two level hierarchical master-slave paradigms.

➢ Receive Update_Solution requests from the worker nodes, and send back the current local best solution to them.

➢ Divide the whole tasks into particles, of which the size could be either fixed or flexible, and record the completion information for every task particles, which includes whether this particle has been computed or not, and ID of the worker node by which this particle is computed.

➢ Receive Ask_For_Tasks requests from the worker nodes, and send back a task particle which has not been computed yet to them.

The required communications between the worker nodes and the master node include:

➢ In a fixed time interval, update the local best solution found by all worker nodes, which is recorded in the master node at any time.

➢ As soon as the worker nodes finish their task particle, send Ask_For_Tasks requests to the master node to ask whether there is any task particle that could be computed by them, and this is decided by the master node and judged based on several factors, which will be discussed thoroughly in section 3.

### III. Load Distribution Strategy

In this section, we consider an existing load distribution strategy, the Adaptive Contracting Within Neighborhood strategy, and our proposed local balancing algorithm, the Proportional Fairness Strategy, with the aim to efficiently balance tasks among the master nodes.

#### A. The Adaptive Contracting Within Neighborhood Strategy (ACMN)

The existing ACWN strategy [3] is derived from the randomized allocation (RAND) [13] strategy, which performs the local balancing only within randomly chosen neighborhood. This is to say that RAND strategy takes no consideration of already allocated burden of nodes. With the ACWN strategy, the supervisor node will always select the least loaded master node in the neighborhood as the receipt of a newly generated subproblem.

### B. The Proportional Fairness Strategy (PFS)

The ACWN strategy works well when the computing power in every node is of similar levels. However, the parallel resource can suffer in a hostile environment where the nodes in the system are in several different conditions, and variety of the network qualities exists. In such a situation, we should not only take account the load which is already allocated in a processing set, but also consider the potency of computing from the view of condition variance.

The Proportional Fairness Strategy differs from the ACWN by taking the computational performance of computing process sets into account. The load balancing strategy manages to assign additional subproblems to the master nodes proportionally to their evaluated performance. The key idea here is that the better the computational performance is, the more subproblems will be allocated. The computational performance of one master node and its processing set is measured according to the following factors: the number of worker nodes in this processing set, and the average execution time of the subproblems which this processing set has already completed. When the supervisor process is ready to reallocate subproblems to one master node, the supervisor process will analyze the computational performance in order to keep the number of assigned subproblems on that master node as follows:

$$N_{subpro(i)} = A_{subpro} \times \frac{T_{subpro(i)} N_{workers(i)}}{\sum_j T_{subpro(j)} N_{workers(j)}}$$

Where $N_{subpro(i)}$, $T_{subpro(i)}$ and $N_{workers(i)}$ denote the number of assigned subproblems, the average subproblem execution time and the number of worker nodes controlled on master $i$ respectively. And $A_{subpro}$ means the number of unexecuted subproblems.

## IV. IMPLEMENTATION

The theoretical research of this paper has been implemented on parallel computing testbeds consisting of a super computer. With the purpose to efficiently improve the computing performance, the communications have been dedicatedly studied. We design a strategy which can greatly reduce the redundant data transferred among processes during communications.

The studied problem is the Flowshop Scheduling Problem. The objective of the problem is to find the best global permutation in the feasible field which minimizes the makespan of all the processed jobs.

### A. The Arrangement of Data Transfer

The bytes exchanged in the application include those from supervisor to masters and from masters to workers, and those in the reverse way. The sizes of data transferred between masters and workers are depended on their purposes and directions. From workers to masters, the messages, in which the newly found local best solutions are contained, are sent out irregularly. This means that the non-blocking detection should keep working on the master's side. When a worker finishes a subproblem and generates a better solution, it makes a connection immediately with the master. In this communication, only the better solution itself is needed to transfer, therefore there are SIZE_OF_INTEGER bytes needed to be transferred in the message except the overhead used as tag head and tail. The SIZE_OF_INTEGER denotes the number of bytes of an integer data type in particular implementation system.

Another communication from masters to workers is to allocate subproblems. In a specific Flowshop Scheduling Problem, the subproblems can be viewed as leaves in a branch and bound tree, and every leaf in the tree denotes a collection of multiple permutations. There will be $n!/(n-k-1)!$ subproblems in the kth floor of the tree, and every subproblem has $(n-k-1)!$ leaves. Since the shape of the tree is only determined by the size of the problem, a subproblem can be traced back as long as its ID number is acknowledged. The ID number of a specific subproblem can be obtained as follows.

$$ID(k,m) = \begin{cases} \sum_{i=1}^{k} \frac{n!}{(n-i)!} + m, k >= 1 \\ m, k = 0 \end{cases}$$

Where $k$ and $m$ denote the level of the floor and the position in that floor of a subproblem, respectively. The following constraints make the equation more sensible.

$$0 \le k \le n-1$$

$$0 \le m \le \frac{n!}{(n-k-1)!} - 1$$

$$0 \le ID(k,m) \le \sum_{i=1}^{n} \frac{n!}{(n-i)!} - 1$$

At the beginning of the application, the benchmarks will be replicated in every worker nodes automatically as part of the source data of the program. Therefore, it is simply required the transmission of the ID numbers of these subproblems, avoiding redundant transfer of the whole problem subsets.

From supervisors to masters, the information transferred is basically for the awareness of the local best solution as well as to achieve load balancing. The following part will give this a carefully consideration.

The load can be reallocated by two very different methods when one master is overloaded. The first is to send back a single higher level subproblem's ID number to the supervisor, who will later reallocate this subproblem to other masters. By doing this, the master loses the whole load of this subproblem, including all the descendant nodes representing lower level subproblems in the bound and branch tree. There is only one ID number needed to be transferred, so the payload of communication is trivial. However, the potential drawback is that since there is only one subproblem being exchanged, the load balancing between masters may not be fluent enough and may lead to slight deficiency.

The second method is to choose multiple lower level subproblems to send. As shown in figure 2, these subproblems can be chosen from several node levels and consequently have

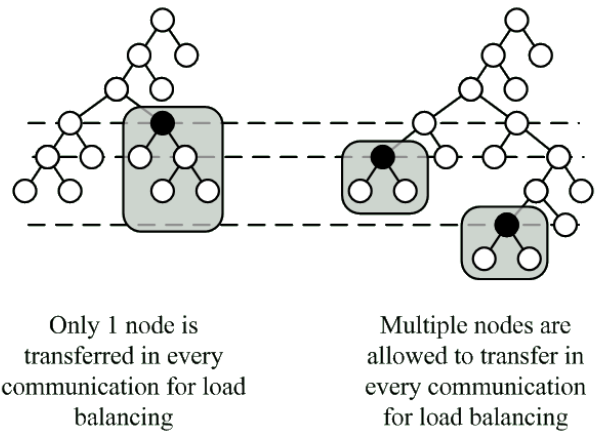

Only 1 node is transferred in every communication for load balancing

Multiple nodes are allowed to transfer in every communication for load balancing the different computational complexities. In one reallocation, there is a group of ID numbers packed together that should be transferred. And there also need a tag indicating how many ID numbers are in the package.

Fig. 2. Two distinctive transferring strategies.

## V. PERFORMANCE EVALUATION

In this sector, we conduct the computational experiments and present the numerical results. We use the enhanced Johnson's algorithm as the sequential method. Although there are many great algorithms proposed for the Flowshop Scheduling Problem, this method is enough for the evaluation of the performance in parallelization. Based on Johnson's algorithm we develop a parallel version which implements the strategies discussed in the previous paragraphs. The target problem is resolved based on the branch and bound algorithm. Meanwhile, the performances of the two proposed distribution strategies, i.e., ACWN and PFS, and that of the load distribution strategy are presented.

The computational resource is listed in table I. The parallel middleware used is MPICH 1.2.7, which is one of the implementation of the Message Passing Interface standard.

Table 1. The computational resource

| number of nodes | specification of a single node | parallel middleware | number of CPUs per node |
|---|---|---|---|
| 21 | Opteron 280 CPU at 2.4GHz | MPICH 1.2.7 | 4 |
| 1 | Opteron 280 CPU at 2.4GHz | MPICH 1.2.7 | 2 |

During the research, we choose a group of benchmarks from the well known benchmarks made available by Taillard [7]. And we also generate a great many benchmarks randomly in order to evaluate the statistic performance of the proposed schemes.

From the comparison it shows that the parallel solution with no internodes communication as well as the static average task allocation always takes a longer computing time than the one with the fixed-interval communication and the dynamic task allocation. This is reasonable, because the internodes communications can share the computation among masters,

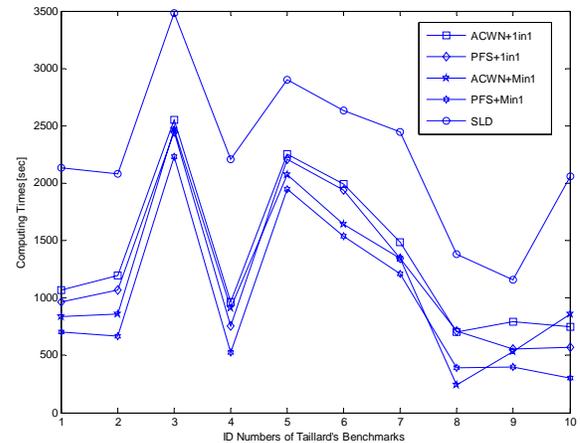

Fig. 3. Efficiency comparison of different strategies dealing with Taillard's benchmarks

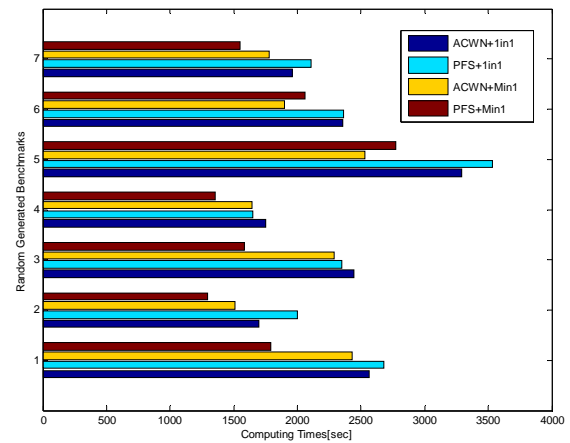

Fig. 4. Efficiency comparison of different strategies dealing with randomly generated benchmarks

and therefore achieve a fairer load distribution.

Figure 3 shows the results of Taillard's benchmarks at a 20 Jobs / 20 Machines complexity. The benchmarks are from #21 to #30 according to sequence provided in [7]. The symbols ACWN, PFS and SLD denote the ACWN strategy, the PFS strategy and the static load distribution strategy respectively. '1in1' means that the case implements the load distribution strategy with the restriction that only one node is transferred in every communication for load balancing, and 'Min1' means that the case allows multiple nodes be transferred in every communication. From the figure, it comes to a clear conclusion that static load distribution loses a great many of efficiency comparing to other schemes, and the performances of the ACWN and the PFS when with 1 node transferred in 1 communication look very similar. This is due to the fact that the parallel computational resource consists of computing nodes in the similar computing standards.

Figure 4 depicts the general results of the randomly generated benchmarks. The processing times are calculated from a normal distribution with a mean value of 50 and a standard deviation of 25. Some other pairs of mean values and standard deviations have also been examined. Since the work focuses on the efficiency of the parallel computing, we consider that it is already adequate to demonstrate the work.

From the figure, although PFS has shorter computing times in some benchmarks, the principle shows that the PFS can achieve better load balancing performance when processing sets have a variety of computing powers. The PFS with multiple nodes transferred in 1 communication for load reallocation achieved an improvement of roughly 13~15% comparing the ACWN.

## VI. Conclusions

In this paper, we provide a parallel implementation of the Flowshop Scheduling Problem, and propose the Proportional Fairness Strategy to keep load balance of a parallel application. We also discuss two different node-transfer schemes of parallel branch and bound architecture. The problem addressed in this paper is the Flowshop Scheduling Problem, which is parallelized using the hierarchical master-worker paradigm. The application is then implemented on a super computer consisting of 86 CPUs using MPICH as the middleware. The experimental results show that by taking computing performance into account, the Proportional Fairness Strategy improves the efficiency of parallelization comparing with the conventional ACWN strategy in terms of computing time by 5~10%. With the multiple nodes transferring in every communication of load balancing, we come to a conclusion that redundant data could be reduced if the ID number, rather than the information of node, is contained in the communication. From the results, it is known that with both these two strategies the computing time can achieve 13~15% improvement.

The experiments behind this paper are done in the LAN scenario. It is expected to examine if the Proportional Fairness Strategy in load distribution will lead to a bigger improvement of performance in WAN, where the network quality is even worse.